\documentclass[12pt]{article}

\usepackage[margin=1in]{geometry}
\usepackage{setspace}
\usepackage{amsmath}
\usepackage{amssymb}
\usepackage{amsthm}
\usepackage{exscale}
\usepackage[mathscr]{eucal}
\usepackage{bm}
\usepackage{eqlist} 
\usepackage[final]{graphicx}
\usepackage[dvipsnames]{color}
\usepackage{verbatim}
\usepackage{natbib}
\usepackage{caption}
\usepackage{enumerate}
\usepackage{subcaption}
\usepackage{algorithm}
\usepackage{algpseudocode}

\usepackage{etoolbox}

\def\be{\begin{equation}}
\def\ee{\end{equation}}
\def\ben{\begin{equation*}}
\def\een{\end{equation*}}
\def\bea{\begin{eqnarray}}
\def\eea{\end{eqnarray}}
\def\bean{\begin{eqnarray*}}
\def\eean{\end{eqnarray*}}
\def\bep{\begin{prop}}
\def\eep{\end{prop}}
\def\bc{\begin{center}}
\def\ec{\end{center}}

\makeatletter
\patchcmd{\env@cases}{1.2}{0.6}{}{}
\makeatother

\DeclareGraphicsExtensions{.pdf, .jpg}
\DeclareMathOperator*{\argmax}{arg\,max}
\DeclareMathOperator*{\argmin}{arg\,min}

\newtheorem{thm}{Theorem}
\newtheorem{lemma}{Lemma}

\theoremstyle{definition}
\newtheorem{definition}{Definition}

\newcommand{\E}{\mathrm{E}}
\newcommand\independent{\protect\mathpalette{\protect\independenT}{\perp}}
\def\independenT#1#2{\mathrel{\rlap{$#1#2$}\mkern2mu{#1#2}}}

\newcommand{\hbs}[1]{\boldsymbol{\hat{#1}}}
\newcommand{\bs}[1]{\boldsymbol{#1}}
\newcommand{\mc}[1]{\mathcal{#1}}

\title{\Large  Nonparametric mixture of Gaussian graphical models}
\author{Kevin Lee and Lingzhou Xue \\ Department of Statistics, The Pennsylvania State University}
\date{December 2015}

\begin{document}

\maketitle

\abstract{
Graphical model has been widely used to investigate the complex dependence structure of high-dimensional data, and it is common to assume that observed data follow a homogeneous graphical model. However, observations usually come from different resources and have heterogeneous hidden commonality in real-world applications. Thus, it is of great importance to estimate heterogeneous dependencies and discover subpopulation with certain commonality across the whole population. In this work, we introduce a novel regularized estimation scheme for learning nonparametric mixture of Gaussian graphical models, which extends the methodology and applicability of Gaussian graphical models and mixture models. We propose a unified penalized likelihood approach to effectively estimate nonparametric functional parameters and heterogeneous graphical parameters. We further design an efficient generalized effective EM algorithm to address three significant challenges: high-dimensionality, non-convexity, and label switching. Theoretically, we study both the algorithmic convergence of our proposed algorithm and the asymptotic properties of our proposed estimators. Numerically, we demonstrate the performance of our method in simulation studies and a real application to estimate human brain functional connectivity from ADHD imaging data, where two heterogeneous conditional dependencies are explained through profiling demographic variables and supported by existing scientific findings.

}


\section{\large Introduction}

Graphical model has been widely used to investigate the complex dependence structure of high-dimensional data, and it has successful applications in various research fields. For example, in bioinformatics, graphical model is used in exploring the patterns of association in gene expression data \citep{dobra-etal-2004,schafer-strimmer-2005}, binary genomic data \citep{wang-etal-2011,xue-etal-2012}, cell signalling data \citep{voorman2014}, among others. Due to advances in functional magnetic resonance imaging (fMRI), investigating brain function connectivity becomes increasingly important \citep{ryali-etal-2012}. Gaussian graphical model has been extensively used in estimating the functional connectivity in brain imaging \citep{ng-etal-2013,varoquaux-etal-2010}. The central question here is to infer conditional dependencies or independencies from high-dimensional fMRI data. {In the current literature, it is common to assume that high-dimensional data come from a homogeneous resource and follow a parametric or semiparametric graphical model, for instance, Gaussian graphical model and its variants \citep{meinshausen-buhlmann-2006,yuan-lin-2007,friedman-etal-2008,peng-etal-2009,witten-etal-2011,cai-etal-2011,liu-etal-2012,xue-zou-2012,venkat-etal-2012,ma-etal-2013,danaher-etal-2014}.}

However, it is very common in real-world applications that observed data come from different resources and may have heterogeneous dependencies across the whole population.
For instance, genetic variations data and gene expression data of the international HapMap project \citep{hapmap-2010} consist of four representative populations in the world. 
Our research is motivated by exploring the heterogeneous dependencies of human brain fMRI data to study the Attention Deficit Hyperactivity Disorder (ADHD). The famous ADHD-200 Global Competition data \citep{biswal-etal-2010} aggregated across 8 independent imaging sites. Thus, it is very important to estimate heterogeneous dependencies and discover subpopulation with certain commonality. In the current literature, there are two different arguments about whether the gender affects the ADHD. On the one hand, some studies show that there is a gender difference in ADHD \citep{sauver-etal-2004}. On the other hand, other studies argue that the ADHD is not systematically different between boys and girls \citep{bauermeister-etal-2007}. It is likely that there may exist two different subpopulations with hidden commonality among the whole population, corresponding to two existing arguments respectively. Section 6 confirms this conjecture. More specifically, we show that both arguments may be explained through investigating heterogeneous brain functional connectivity using our proposed method.


In this work, we introduce a novel regularized estimation scheme for learning nonparametric mixture of Gaussian graphical models, which explores the heterogeneous dependencies of high-dimensional data. Denote by $\mc{G}^z(\bold{x})=(\mc{V},\mc{E}^z)$ the graphical model of random vector $\bold{x}\in\mathbb{R}^p$ with a univariate covariate $Z=z$, vertex set $\mc{V}=\{1,\ldots,p\}$ and edge set $\mc{E}^z$, where edge set $\mc{E}^z$ may depend on $z$. Let $K$ be number of mixtures. Let $\mc{G}^z_k(\bold{x})=(\mc{V},\mc{E}^z_k)$ represent the Gaussian graphical model in the $k$-th subpopulation. Define $\mc{C}$ as the latent class variable satisfying that $P(\mc{C}=k|Z=z)=\pi_k(z)$, where $\pi_k(z)$ is a nonparametric mixing proportion function. Throughout this paper, we consider the following nonparametric mixture of Gaussian graphical models given some covariate $Z=z$:
\be\label{mixtureGGM}
\mc{G}^z(\bold{x}) =\pi_1(z)\mc{G}^z_1(\bold{x}) +\pi_2(z)\mc{G}^z_2(\bold{x})+\cdots+\pi_K(z)\mc{G}^z_K(\bold{x}),
\ee
where $\pi_k(\cdot)$'s are nonparametric mixing proportion functions, and $\pi_1(z)+\ldots+\pi_K(z)=1$ for any $z\in\mathbb{R}$. Mixture models are powerful to effectively identify subpopulations with hidden commonality within the whole population \citep{lindsay-1995,mclachlan2004finite}. Mixture models were extensively studied in the classical low-dimensional scenarios, but receive less attention in high-dimensional statistical learning. Recently, \cite{stadler-etal-2010} studied $\ell_1$ penalization for mixtures of high-dimensional regression models, and \cite{ruan-etal-2011} studied mixtures of Gaussian graphical models. However, it is more challenging to learn nonparametric mixture of Gaussian graphical models \eqref{mixtureGGM} in the presence of high-dimensionality, non-convexity and label switching. In Section 2, we propose a unified penalized likelihood approach to effectively estimate both nonparametric functional parameters and heterogeneous graphical parameters. To estimate nonparametric functional parameters, we adopt the idea of kernel regression technique and employ a local likelihood approach. Section 3 designs an efficient generalized effective EM algorithm to address three aforementioned challenges simultaneously. Furthermore, we propose an effective criterion to choose the number of mixtures, tuning parameter and bandwidth. We study the algorithmic convergence of our EM algorithm and the asymptotic result of our estimates in Section 4. Simulation studies and the real application to time-varying brain functional connectivity estimation are demonstrated in Sections 5 and 6 respectively. In Section 6, we discover two heterogeneous dependencies in the ADHD brain functional connectivity, which are explained through profiling demographic variables and supported by existing scientific findings. Our results provide helpful insights to study two different ADHD subpopulations with hidden commonality. 

\section{\large Nonparametric mixture of Gaussian graphical models}


Given some univariate covariate $Z=z$, nonparametric mixture \eqref{mixtureGGM} entails that $\bold{X}=\bold{x}$ follows a nonparametric finite mixture of multivariate Gaussian distributions:
\begin{equation}\label{model}
\bold{X}=\bold{x}\left.\right|_{Z=z} \sim_d \sum_{k=1}^{K}\pi_{k}(z)N_p(\boldsymbol{\mu}_k(z), \boldsymbol{\Sigma}_k(z)),
\end{equation}
where $\mc{G}^z_k(\bold{x})$ corresponds to a multivariate normal distribution $N_p(\boldsymbol{\mu}_k(z), \boldsymbol{\Sigma}_k(z))$ for $k=1,\ldots,K$. Let $\boldsymbol{\Theta}_k(z)=(\theta_{kij}(z))_{p\times p}$ be the precision matrix in the $k$-th mixture. Then, $\theta_{kij}(z)$'s specify the graphical model $\mc{G}^z_k(\bold{x})$ in the $k$-th mixture \citep{dempster-1972}. Specifically, given $Z=z$, zeroes in $\theta_{kij}(z)$'s are equivalent to conditional independencies of $\bold{X}$ in the $k$-th mixture. Thus, zeroes in $\theta_{kij}(z)$'s can be translated to a meaningful graphical model:
$$
\theta_{kij}(z)\neq 0 \ \Longleftrightarrow \ X_i \not \independent X_j \left.\right|_{\bold{X} \setminus \{X_i,X_j\},\mc{C}=k,Z=z}  \ \Longleftrightarrow \ \{i,j\}\in\mc{E}^z_k.
$$
For ease of presentation, we start with the known a priori number of mixtures $K$, and we present an effective information criterion to select $K$ in Section 3.2.

Given the independent data $\{(\bold{x}_{n}, z_{n}), n=1, \ldots, N\}$, our goal is to estimate nonparametric functions $\pi_k(\cdot)$'s and functional parameters $\boldsymbol{\mu}_k(\cdot)$'s and $\boldsymbol{\Theta}_k(\cdot)$'s. The average log-likelihood function for the observed data is given by
$$
\ell_N = \frac{1}N\sum_{n=1}^{N}\log\left[\sum_{k=1}^{K}\pi_{k}(z_{n})\phi(\bold{x}_{n}|\boldsymbol{\mu}_k(z_{n}), \boldsymbol{\Theta}_k(z_{n}))\right].
$$
where $\phi(\cdot|\boldsymbol{\mu}, \boldsymbol{\Theta})$ is the density of a multivariate Gaussian distribution $N_p(\boldsymbol{\mu}, \boldsymbol{\Theta}^{-1})$.

In view of functional parameters, we employ kernel regression techniques to estimate $\pi_k(z)$'s, $\boldsymbol{\mu}_k(z)$'s, and $\boldsymbol{\Theta}_k(z)$'s for any $z\in\mathbb{R}$. To this end, we define the local average log-likelihood function as
$$
\ell_z \ = \ \frac{1}N\sum_{n=1}^{N}\log\left[\sum_{k=1}^{K}\pi^z_{k}\phi(\bold{x}_{n}|\boldsymbol{\mu}_k^{z}, \boldsymbol{\Theta}_k^{z})\right]K_{h}(z_n-z),
$$
where $K_h(\cdot) = h^{-1}K(\cdot / h)$ is a symmetric kernel function with bandwidth $h$. Given the local average log-likelihood $\ell_z$, we maximize the $\ell_1$-penalized local log-likelihood to estimate local constants $\pi_{k}^z$'s, local vectors $\boldsymbol{\mu}_{k}^z$'s, and local constant matrices $\boldsymbol{\Theta}^z_k$'s as follows:
\begin{equation}\label{opti}
\max_{\{(\pi_{k}^z,\bs{\mu}^z_k,\bs{\Theta}^z_k)\}_{k=1,\ldots,K}} \mathcal{L}_z
\quad := \quad
\max_{\{(\pi_{k}^z,\bs{\mu}^z_k,\bs{\Theta}^z_k)\}_{k=1,\ldots,K}}  \ell_z - \lambda\sum_{k=1}^{K}\|\boldsymbol{\Theta}^z_k\|_{1,\text{off}} ,
\end{equation}
where $\|\cdot\|_{1,\text{off}}$ is the entrywise matrix $\ell_{1}$ norm of the off-diagonal elements. Instead of $\ell_1$ penalization, we may also consider the folded concave penalized estimation to encourage sparsity in $\boldsymbol{\Theta}^z_k$'s \citep{fan-etal-2009,fan-etal-2014}. For space consideration, we only focus on $\ell_1$ penalization in this work.

\medskip
\textbf{Remark 1}: when $K=1$, nonparametric mixture of graphical models \eqref{mixtureGGM} reduces to covariate-dependent graphical model. When the covariate $Z$ represents the varying time, $\mc{G}^z(\bold{x})$ becomes time-varying graphical model \citep{ahmed-xing-2009,kolar-song-ahmed-xing-2010,zhou-lafferty-wasserman-2010}, which is solved by using the penalized likelihood approach:
$$
\hbs{\Theta}^z= \argmax_{\bs{\Theta}^z} \ \frac{1}N\sum_{n=1}^{N}\log\left[\phi(\bold{x}_{n}|\boldsymbol{\mu}^{z}, \boldsymbol{\Theta}^{z})\right]K_h(z_n-z) - \lambda\|\boldsymbol{\Theta}^z\|_{1,\text{off}}.
$$

\medskip
\textbf{Remark 2}: when $\mc{E}_k$ does not depend on covariate $z$, nonparametric mixture \eqref{mixtureGGM} reduces to semiparametric mixture of graphical models. Let $\mc{G}_k(\bold{x})=(\mc{V},\mc{E}_k)$ be the $k$-th mixture. The semiparametric mixture becomes
$
\mc{G}^z(\bold{x}) =\pi_1(z)\mc{G}_1(\bold{x}) +\cdots+\pi_K(z)\mc{G}_K(\bold{x}),
$
where $\pi_1(z)+\ldots+\pi_K(z)=1$ for any $z\in\mathbb{R}$. Conditioning on $Z=z$, $\bold{x}$ follows a semiparametric finite mixture
$
\sum_{k=1}^{K}\pi_{k}(z)N_p(\boldsymbol{\mu}_k, \boldsymbol{\Sigma}_k).
$
Now we introduce local constants $\pi_{k}^z$'s to approximate $\pi_{k}(z)$'s. Next, we may solve local estimates and global estimates of $\boldsymbol{\Theta}_k$'s. Firstly, we solve local constants $\tilde \pi_{k}^z$'s from
\begin{equation*}
\max_{\{(\pi_{k}^z,\bs{\Theta}_k)\}_{k=1,\ldots,K}} \ \frac{1}N\sum_{n=1}^{N}\mbox{log}\left[\sum_{k=1}^{K}\pi_{k}^z \phi(\bold{x}_{n}|\boldsymbol{\mu}_k, \boldsymbol{\Theta}_k)\right] K_h(z_n-z) - \lambda\sum_{k=1}^{K}\|\boldsymbol{\Theta}_k\|_{1,\text{off}}.
\end{equation*}
After obtaining local estimates $\tilde \pi_{k}(z_n)$'s for $n=1,\ldots,N$, we solve global estimates via
\begin{equation*}
\max_{\{\bs{\Theta}_k\}_{k=1,\ldots,K}} \ \frac{1}N\sum_{n=1}^{N}\mbox{log}\left[\sum_{k=1}^{K}\tilde \pi_{k}(z_{n})\phi(\bold{x}_{n}|\boldsymbol{\mu}_k, \boldsymbol{\Theta}_k)\right] - \lambda\sum_{k=1}^{K}\|\boldsymbol{\Theta}_k\|_{1,\text{off}}.
\end{equation*}

\medskip
\textbf{Remark 3}: when there is no covariate, nonparametric mixture \eqref{mixtureGGM} further reduces to finite mixture of Gaussian graphical models \citep{ruan-etal-2011}, i.e.
$$
\mc{G}(\bold{x}) =\pi_1\mc{G}_1(\bold{x}) +\pi_2\mc{G}_2(\bold{x})+\cdots+\pi_K\mc{G}_K(\bold{x}),
$$
that is, $\bold{x} \sim_d \sum_{k=1}^{K}\pi_{k}N_p(\boldsymbol{\mu}_k, \boldsymbol{\Sigma}_k)$, where $\pi_1+\ldots+\pi_K=1$. This can be solved by
\begin{equation*}
\max_{\{(\pi_{k},\bs{\Theta}_k)\}_{k=1,\ldots,K}} \ \frac{1}N\sum_{n=1}^{N}\log\left[\sum_{k=1}^{K}\pi_{k}\phi(\bold{x}_{n}|\boldsymbol{\mu}_k, \boldsymbol{\Theta}_k)\right] - \lambda\sum_{k=1}^{K}\|\boldsymbol{\Theta}_k\|_{1,\text{off}}.
\end{equation*}

\section{\large Computation}

Expectation-Maximization (EM) algorithm provides a powerful tool to solve latent variable problems in mixture model. \cite{wu-1983} established some general convergence properties, and \cite{balakrishnan-etal-2014} recently studied statistical guarantees on both population level and sample level. However, we need to address two significant challenges when solving the nonparametric mixture \eqref{opti}: 1) non-convex optimization in high dimensions; 2) label switching issue at different grid points of covariate $z$. This section presents a generalized effective EM algorithm to address both challenges, which enjoys some appealing convergence properties as shown in Section 4.

\subsection{Proposed EM algorithm}

Following the spirit of EM algorithm, we view the collected data ${(\bold{x}_{n}, z_n), n=1,\ldots,N}$ to be incomplete, and then define random variables $\boldsymbol{\tau}_n = (\tau_{1n}, \ldots, \tau_{Kn})'$ with
\begin{equation*}
\tau_{kn} = \begin{cases}
1 & \quad \text{if $(\bold{x}_{n}, z_n)$ is in the k-th mixture}, \\[1em]
0 & \quad \text{otherwise}.
\end{cases}
\end{equation*}
to identify the mixture of $(\bold{x}_{n}, z_n)$. Given the complete data $\{(\bold{x}_{n}, z_n, \boldsymbol{\tau}_{n}), n=1, \ldots, N\}$, the complete log-likelihood function is written as
$$
\ell^{\text{cmp}}_{N} = \frac{1}N\sum_{n=1}^{N}\sum_{k=1}^{K}\tau_{kn}\left[\log\pi_{k}(z_{n})+\log\phi(\bold{x}_{n}|\boldsymbol{\mu}_k(z_{n}), \boldsymbol{\Theta}_k(z_{n}))\right].
$$

\begin{figure}[!htb]
\caption{Solution paths for two mixtures at six different grid points.}
\label{fig:Lasso}
  \centering
	\includegraphics[scale=0.5]{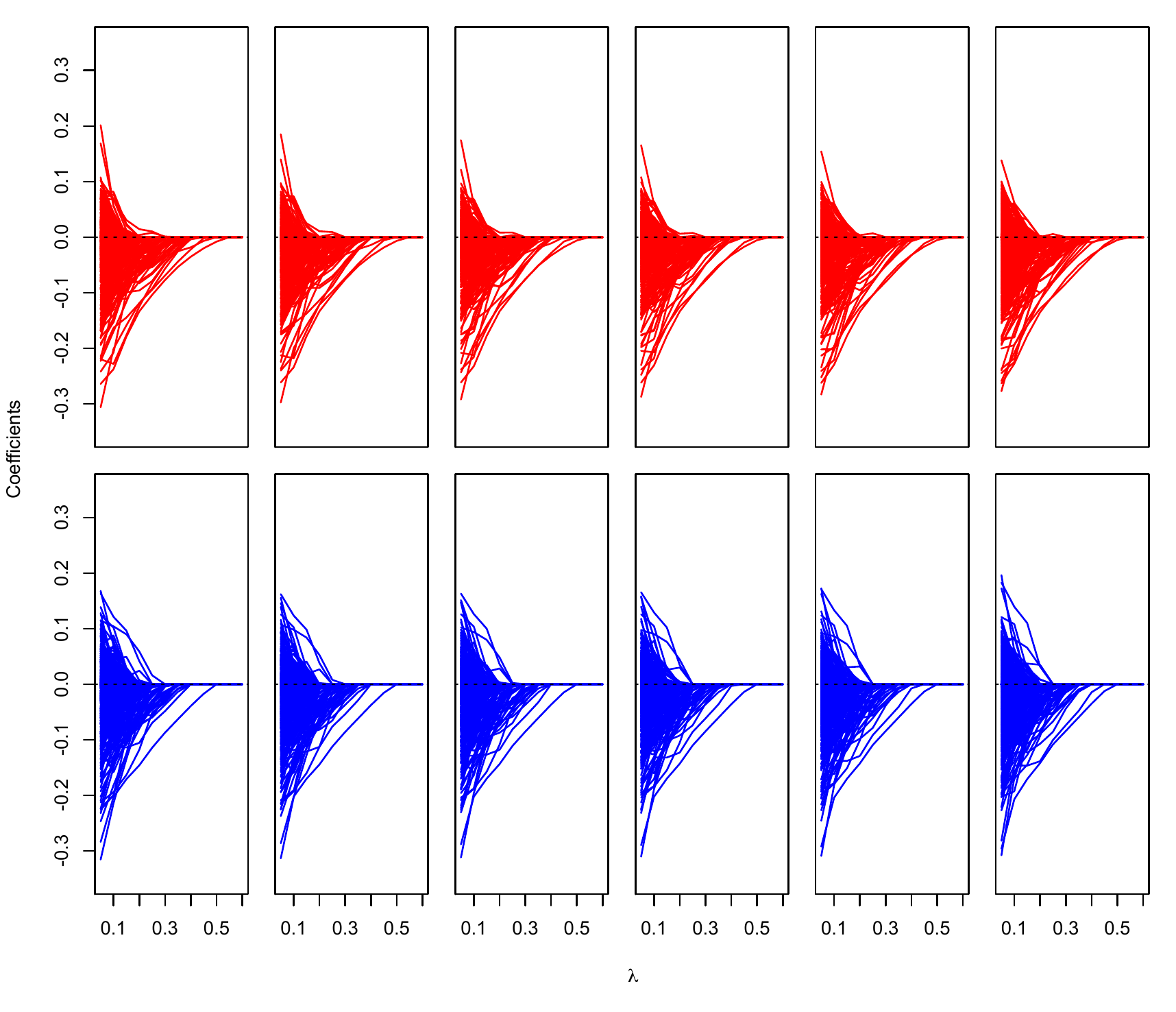}
\end{figure}

Let $z \in \{u_1, \ldots, u_G\}$, the set of grid points. We employ kernel regression techniques to estimate $\pi_{k}(z_{n})$, $\boldsymbol{\mu}_{k}(z_{n})$, and $\boldsymbol{\Theta}_{k}(z_{n})$. Define a local complete log-likelihood as
$$
\ell^{\text{cmp}}_{z} = \frac{1}N\sum_{n=1}^{N}\sum_{k=1}^{K}\tau_{kn}\left[\log\pi_{k}^{z}+\log\phi(\bold{x}_{n}|\boldsymbol{\mu}_k^{z}, \boldsymbol{\Theta}_k^{z})\right] K_{h}(z_{n}-z).
$$
Next, we define the $\ell_{1}$-penalized local log-likelihood function for the complete data as
$$
\mathcal{L}^{\text{cmp}}_{z} \ = \ \ell^{\text{cmp}}_{z}  - \lambda\sum_{k=1}^{K}\|\boldsymbol{\Theta}_{k}^{z}\|_{1,\text{off}}.
$$
Notice that there are potential label switching issues at any two different grid points $z,z' \in \mathcal{U}=\{u_1, \ldots, u_G\}$. To solve this issue, we propose the following generalized effective EM algorithm. Given the current estimates of parameters, E-step estimates all conditional expectations \eqref{gamma} at observed $\{z_1, \ldots, z_N\}$. M-step uses the obtained common conditional expectations to update all estimates of parameters at each grid point in $\mathcal{U}$. Hence, we effectively prevent the label switching issue at different grid points. As shown in Figure \ref{fig:Lasso}, two solution paths from simulation studies in Section 5 demonstrate that two mixtures are consistently identified at different grid points.

In what follows, we present algorithm details. Since only $(\bold{x}_{n}, z_n)$'s are observed, we treat $\boldsymbol{\tau}_{n}$'s as missing data. In the $(t+1)$-th iteration, $t=0,1,2,\ldots$, E-step employs the $t$-th iterated solution $\pi_{k}^{(t)}(z_{n})$, $\boldsymbol{\mu}_{k}^{(t)}(z_{n})$ and $\boldsymbol{\Theta}_{k}^{(t)}(z_{n})$ to compute the conditional expectation of $\tau_{kn}$ given the current estimates. By using Bayes' rule, we have
\begin{equation}\label{gamma}
\gamma_{kn}^{(t+1)} = \frac{\pi_{k}^{(t)}(z_{n})\phi(\bold{x}_{n}|\boldsymbol{\mu}_{k}^{(t)}(z_{n}), \boldsymbol{\Theta}_{k}^{(t)}(z_{n}))}{\sum\limits_{l=1}^{K}\pi_{l}^{(t)}(z_{n})\phi(\bold{x}_{n}|\boldsymbol{\mu}_{l}^{(t)}(z_{n}), \boldsymbol{\Theta}_{l}^{(t)}(z_{n}))}.
\end{equation}
Next, M-step obtains the estimates of parameters from maximizing
$$
\frac{1}N\sum_{n=1}^{N}\sum_{k=1}^{K}\gamma_{kn}^{(t+1)}\left[\log\pi_{k}^{z} + \log \phi(\bold{x}_{n}|\boldsymbol{\mu}_k^{z}, \boldsymbol{\Theta}_k^{z})\right]K_{h}(z_n-z) - \lambda\sum_{k=1}^{K}\|\boldsymbol{\Theta}_k^{z}\|_{1,\text{off}}
$$
subject to the constraint that $\sum_{k=1}^{K}\pi_{k}^{z}=1$. It is equivalent to maximizing
\begin{equation}\label{pi.opt}
\frac{1}N\sum_{n=1}^{N}\sum_{k=1}^{K}\gamma_{kn}^{(t+1)}\log\pi_{k}^{z} K_{h}(z_n-z),
\end{equation}
subject to $\sum_{k=1}^{K}\pi_{k}^{z}=1$, and for $k =1, \ldots, K$, maximizing
\begin{equation}\label{theta.opt}
\frac{1}N\sum_{n=1}^{N}\gamma_{kn}^{(t+1)}\log \phi(\bold{x}_{n}|\boldsymbol{\mu}_k^{z}, \boldsymbol{\Theta}_k^{z}) K_{h}(z_n-z) - \lambda\|\boldsymbol{\Theta}_k^{z}\|_{1,\text{off}},
\end{equation}
respectively. To solve the subproblem \eqref{pi.opt}, we introduce Lagrange multiplier $\alpha$ with the constraint $\sum_{k=1}^{K}\pi_{k}^{z} = 1$. Then in the $(t+1)$-th cycle, for $z \in \{u_{g}, g=1,\ldots,G\}$ we update $\pi_k^z$ by
\begin{equation}\label{pi}
\pi_{k}^{z(t+1)} = \sum_{n=1}^{N}\frac{\gamma_{kn}^{(t+1)}K_{h}(z_n-z)}{\sum_{n'=1}^{N}K_{h}(z_{n'}-z)}.
\end{equation}
In order to solve the subproblem \eqref{theta.opt}, we first simplify \eqref{theta.opt} as,
$$
\frac{1}N\sum_{n=1}^{N}\gamma_{kn}^{(t+1)}\left[\frac{1}{2}\log|\boldsymbol{\Theta}_k^{z}| - \frac{1}{2}(\bold{x}_{n}-\boldsymbol{\mu}_k^{z})'\boldsymbol{\Theta}_k^{z}(\bold{x}_{n}-\boldsymbol{\mu}_k^{z})\right]K_{h}(z_n-z) - \lambda\|\boldsymbol{\Theta}_{k}^{z}\|_{1,\text{off}}.
$$
Then, it is easy to obtain the closed-form update for $\boldsymbol{\mu}_{k}^{z}$, that is
\begin{equation}\label{mu}
\boldsymbol{\mu}_{k}^{z(t+1)} = \sum_{n=1}^{N}\frac{\gamma_{kn}^{(t+1)}K_{h}(z_n-z)\bold{x}_{n}}{\sum_{n'=1}^{N}\gamma_{kn'}^{(t+1)}K_{h}(z_{n'}-z)}.
\end{equation}
Next, we employ the state-of-art optimization algorithm such as \cite{friedman-etal-2008}, \cite{,witten-etal-2011} and \cite{goldfarb-etal-2013} and to solve $\boldsymbol{\Theta}_{k}^{z}$ from
\begin{equation}\label{Sigma}
\boldsymbol{\Theta}_{k}^{z(t+1)} = \argmax_{\boldsymbol{\Theta}_{k}^{z}}\left\{\log|\boldsymbol{\Theta}_k^{z}| - \mbox{tr}(\boldsymbol{\Theta}_{k}^{z}\bold{A}_{k}^{z(t+1)}) - \lambda\|\boldsymbol{\Theta}_k^{z}\|_{1,\text{off}}\right\},
\end{equation}
where
$
\bold{A}_{k}^{z(t+1)} = \sum_{n=1}^{N}\frac{\gamma_{kn}^{(t+1)}K_{h}(z_n-z)}{\sum_{n'=1}^{N}\gamma_{kn'}^{(t+1)}K_{h}(z_{n'}-z)}(\bold{x}_{n}-\boldsymbol{\mu}_{k}^{z(t+1)})(\bold{x}_{n}-\boldsymbol{\mu}_{k}^{z(t+1)})'.
$
Furthermore, we update $\pi_{k}^{(t+1)}(z_{n})$, $\boldsymbol{\mu}_{k}^{(t+1)}(z_{n})$, and $\boldsymbol{\Theta}_{k}^{(t+1)}(z_{n})$, $n=1,\ldots,N$ by linear interpolating $\pi_{k}^{u_{g}(t+1)}$, $\boldsymbol{\mu}_{k}^{u_{g}(t+1)}$, and $\boldsymbol{\Theta}_{k}^{u_{g}(t+1)}$, $g=1,\ldots,G$ respectively. 

\medskip
Now, we summarize the details of our proposed algorithm in Algorithm \ref{al1}.
\begin{algorithm}[!h]
\caption{{Proposed generalized effective EM algorithm}}\label{al1}
\begin{itemize}\itemsep.15in
  \item Initialization of $\pi_{k}^{(0)}(z_{n})$, $\boldsymbol{\mu}_{k}^{(0)}(z_{n})$, and $\boldsymbol{\Theta}_{k}^{(0)}(z_{n})$ for all $k$.
  \item Iteratively solve E-step and M-step with $t=0, 1, 2, \ldots$ till convergence:
  \medskip
  \begin{itemize}\itemsep.15in
    \item \textbf{E-step}: compute $\gamma_{kn}^{(t+1)}$ from \eqref{gamma} for all $k$ and $n$
    \item \textbf{M-step}: compute $\pi_{k}^{z(t+1)}$,  $\boldsymbol{\mu}_{k}^{z(t+1)}$, $\boldsymbol{\Theta}_{k}^{z(t+1)}$ from \eqref{pi}--\eqref{Sigma} for all $k$ and $z \in \mathcal{U}$
  \end{itemize}
\end{itemize}
\end{algorithm}


\subsection{Selection of tuning parameters}

We need to select three tuning parameters: number of mixtures $K$, penalization parameter $\lambda$, and bandwidth $h$. To determine them, we consider the information criterion approach. Bayesian Information Criterion (BIC) has the general form of
$
-2\mathcal{L} + \delta \times df,
$
where $\mathcal{L}$ is the maximum log-likelihood, $\delta = \log N$, and $df$ is the degree of freedom to measure model complexity. To specify the degree of freedom in nonparametric mixture \eqref{model}, we follow \cite{fan-etal-2001} and \cite{huang-etal-2013} to derive the degree of freedom. Denote by
$
\mbox{df} = \tau_{K}h^{-1}|\mathcal{Z}|\left(K(0) - \frac{1}{2}\int K^{2}(t)\,dt\right)
$
the degree of freedom of a univariate nonparametric function, where $\mathcal{Z}$ is the support of the covariate $Z$, and
$
\tau_{K} = \frac{K(0) - \frac{1}{2}\int K^{2}(t)\,dt}{\int\left(K(t) - \frac{1}{2} K * K(t)\right)^{2}\,dt}.
$ Hence, for each pair of $(K, \lambda, h)$, the BIC score is defined as
$$
\mbox{BIC}(K, \lambda, h) =  -2\mathcal{L}  + \log(N) \times df(K, \lambda, h),
$$
where
$$
df(K, \lambda, h) = \left[(K - 1) + \frac{1}{G}\sum_{z \in \mathcal{U}}\left(Kp+\sum_{k=1}^{K}\sum_{i \leq j}I_{\left\{\hat{\Theta}^{z}_{ijk} \neq 0 \right\}} \right) \right] \times \mbox{df}.
$$

We first select $K$ and $\lambda$ by minimizing the BIC score, and then choose $h$ by cross validation (CV). We choose $K$ by minimizing the best available BIC score for each choice of $K$ over different choices of $\lambda$ and $h$. Namely,
$$
\hat K= \argmin_{(\lambda,h)}\mbox{BIC}(K, \lambda, h).
$$
After fixing $K=\hat K$, we choose $\lambda$ by minimizing the best available BIC score for each $\lambda$ over different choices of $h$. Namely,
$$
\hat \lambda= \argmin_{h}\mbox{BIC}(\hat K, \lambda, h).
$$
Lastly, we use the log-likelihood to construct the CV loss, and choose $h$ by CV.






\section{\large Theoretical properties}

This section will first establish the algorithmic convergence of our proposed algorithm, and then prove the asymptotic properties of our proposed estimator. The proofs for following Theorems are specified in the longer version \cite{lee-xue-2015}.

\subsection{Algorithmic convergence}

We first show that our proposed generalized effective EM algorithm (i.e. Algorithm 1) preserves the nice ascent property as the classical EM algorithm with probability tending to 1. Recall that $\mathcal{L}_z(\boldsymbol{\pi}^z,\bs{\mu}^z,\bs{\Theta}^z)=\ell_z(\boldsymbol{\pi}^z,\bs{\mu}^z,\bs{\Theta}^z) - \lambda\sum_{k=1}^{K}\|\boldsymbol{\Theta}^z_k\|_{1,\text{off}}$ is the objective function.  Let $\left\{(\boldsymbol{\pi}^{z(t)}, \boldsymbol{\mu}^{z(t)}, \boldsymbol{\Theta}^{z(t)}): \ t=0,1,2,...\right\}$ be the sequence generated by Algorithm 1.

\begin{thm}
Suppose $h \rightarrow 0$ and $Nh \rightarrow \infty$ as $N \rightarrow \infty$. For any given point $z$ and $t=0,1,2,...$, with probability tending to 1, we always have
\begin{equation*}
\mathcal{L}_z(\boldsymbol{\pi}^{z(t+1)}, \boldsymbol{\mu}^{z(t+1)}, \boldsymbol{\Theta}^{z(t+1)}) \geq \mathcal{L}_z(\boldsymbol{\pi}^{z(t)}, \boldsymbol{\mu}^{z(t)}, \boldsymbol{\Theta}^{z(t)}).
\end{equation*}
\end{thm}

\medskip
Since the objective function \eqref{opti} is non-concave, we focus on the local convergence. Given the ascent property in Theorem 1, we are ready to prove the local convergence result of our proposed EM algorithm in the following theorem. Theorem 2 extends the local convergence result of \cite{stadler-etal-2010} to the nonparametric mixture \eqref{mixtureGGM}.

\begin{thm}
Under the same conditions of Theorem 1, with probability tending to 1, our proposed generalized effective EM algorithm (i.e. Algorithm 1) achieves the local convergence. More specifically, for any given point $z$, every accumulation point $(\bar{\boldsymbol{\pi}}^{z}, \bar{\boldsymbol{\mu}}^{z}, \bar{\boldsymbol{\Theta}}^{z})$ in the sequence $\{(\boldsymbol{\pi}^{z(t)}, \boldsymbol{\mu}^{z(t)}, \boldsymbol{\Theta}^{z(t)}): \ t=0,1,2,\ldots\}$ is a stationary point of the objective function $\mathcal{L}_{z}(\boldsymbol{\pi}^z,\bs{\mu}^z,\bs{\Theta}^z)$ in \eqref{opti} with probability tending to 1.
\end{thm}

\bigskip
\textbf{Remark 4}: when learning the nonparametric mixture of Gaussian graphical models \eqref{mixtureGGM}, Theorems 1-2 prove that both the ascent property and the local convergence hold for Algorithm 1 with probability tending to 1. It is obvious that Theorems 1-2 can be easily extended to the semiparametric mixture $\mc{G}^z(\bold{x}) =\pi_1(z)\mc{G}_1(\bold{x}) +\cdots+\pi_K(z)\mc{G}_K(\bold{x})$ in Remark 2. When there is no covariate $z$, we may further extend Theorems 1-2 and obtain the exact ascent property that $\mathcal{L}(\boldsymbol{\pi}^{(t+1)}, \boldsymbol{\mu}^{(t+1)}, \boldsymbol{\Theta}^{(t+1)}) \geq \mathcal{L}(\boldsymbol{\pi}^{(t)}, \boldsymbol{\mu}^{(t)}, \boldsymbol{\Theta}^{(t)})$ and the exact local convergence for the finite mixture of Gaussian graphical models $\mc{G}(\bold{x}) =\pi_1\mc{G}_1(\bold{x}) +\cdots+\pi_K\mc{G}_K(\bold{x})$ \citep{ruan-etal-2011} in Remark 3.

\subsection{Asymptotic properties}

Let $\boldsymbol{\omega}(z) = (\boldsymbol{\pi}(z), \boldsymbol{\mu}(z), \boldsymbol{\Theta}(z))$ be the true functional parameters in model \eqref{model}. Define $\text{vec}(\cdot)$ as the vectorization of a matrix. We introduce some regularity conditions.

\begin{enumerate}[A.]
\item $\{(\bold{x}_{n}, z_{n}), n=1,\ldots,N\}$ are independent and identically distributed as $(\bold{X}, Z)$. The support for $Z$, denoted by $\mathcal{Z}$, is compact subset of $\mathbb{R}^{1}$.
\item $\boldsymbol{\omega}(z)$ have continuous second derivatives, and $\pi_{k}(z)>0$ for any $z \in \mathcal{Z}$.
\item The marginal density $f(z)$ of $Z$ is twice continuously differentiable and positive.
\item The kernel function $K(\cdot)$ is symmetric, continuous, and has a closed and bounded support and satisfy following conditions:
\begin{equation*}
\begin{split}
& \int K(u)\,du = 1, \hspace{1em} \int uK(u)\,du = 0, \hspace{1em} \int u^{2}K(u)\,du < \infty, \\
& \int K^{2}(u)\,du < \infty, \hspace{1em} \int |K^{3}(u)|\,du < \infty.
\end{split}
\end{equation*}
\item There exists a function $M(\bold{x})$ with $\E[M(\bold{X})] < \infty$, such that for all $\bold{x}$ and all $\boldsymbol{\omega}^{z}$ in a neighborhood of $\boldsymbol{\omega}(z)$, $|\partial^{3}\ell(\boldsymbol{\omega}(z),\bold{x})/\partial\omega_{i}\partial\omega_{j}\partial\omega_{l}| < M(\bold{x})$ holds.
\item $
\E\left(|\frac{\partial\ell(\boldsymbol{\omega}(z),\bold{X})}{\partial\omega_{i}}|^{3}\right) < \infty$ and $\E\left(\{\frac{\partial^{2}\ell(\boldsymbol{\omega}(z),\bold{X})}{\partial\omega_{i}\partial\omega_{j}}\}^{2}\right) < \infty$ hold for all $i$ and $j$
\item $h \rightarrow 0$, $Nh \rightarrow \infty$ and $Nh^{5} = O(1)$ as $N \rightarrow \infty$.
\end{enumerate}

Conditions A.-G. are standard and have been used in \cite{mack-silverman-1982}, \cite{zhou-lafferty-wasserman-2010}, \cite{huang-etal-2013}, and many others.

In the following theorem, we prove the asymptotic properties of the local solution $(\hat{\boldsymbol{\pi}}(z),\hat{\boldsymbol{\mu}}(z),\hat{\boldsymbol{\Theta}}(z))$ for the nonparametric mixture \eqref{model}.

\begin{thm}\label{theorem3}
If $\lambda = O((Nh)^{-1/2})$ as $N \rightarrow \infty$,  under Conditions A.-G., there exists a local maximizer $(\hat{\boldsymbol{\pi}}(z),\hat{\boldsymbol{\mu}}(z),\hat{\boldsymbol{\Theta}}(z))$ of \eqref{opti} such that
$$
\sqrt{Nh}(\hat{\boldsymbol{\pi}}(z) - \boldsymbol{\pi}(z)) = O_{p}(1),
\quad \quad
\sqrt{Nh}(\hat{\boldsymbol{\mu}}(z) - \boldsymbol{\mu}(z)) = O_{p}(1),
$$
and
$$
\sqrt{Nh}(\text{vec}(\hat{\boldsymbol{\Theta}}(z)) - \text{vec}(\boldsymbol{\Theta}(z))) = O_{p}(1).
$$
\end{thm}

\bigskip
\textbf{Remark 5}: In view of Theorem \ref{theorem3}, our proposed method delivers a nice local solution to well estimate both the nonparametric mixing proportions $\boldsymbol{\pi}(z)$ and the heterogenous graphical parameters $\boldsymbol{\mu}(z)$ and $\boldsymbol{\Theta}(z)$. In particular, when the condition that $\min_k\min_{(i,j)\in\mc{E}^z_k} |\theta_{kij}(z)|\gg (Nh)^{-1/2}$ is satisfied, the local maximizer $(\hat{\boldsymbol{\pi}}(z),\hat{\boldsymbol{\mu}}(z),\hat{\boldsymbol{\Theta}}(z))$ of \eqref{opti} would recover all true edges in nonparametric mixture \eqref{mixtureGGM}, that is, $|\hat\theta_{kij}(z)|>0$ for any $k$ and $(i,j)\in\mc{E}^z_k$.

{\section{\large Simulation Studies}}

In this section, we conduct two simulation studies. Section 5.1 considers a mixture of AR and block diagonal dependencies, while Section 5.2 considers a mixture of AR and random sparse dependencies. Before proceeding, we first introduce the common simulation setting and several measurements to compare numerical performances.

We consider the following mixing proportion functions with $K=2$ and $\mathcal{Z}=[0,1]$,
$$
\pi_{1}(z) = \frac{\exp(0.5z)}{1+\exp(0.5z)} \quad \mbox{and} \quad \pi_{2}(z) = 1 - \pi_{1}(z).
$$
We assume that data are observed at equally spaced points in $\mathcal{Z}$, and 50 observations were generated for each point. We specify dimension $p$ as 50 or 100.
To assess numerical performance, we introduce several average metrics over $100$ replications:

\medskip
\begin{itemize}\itemsep .1in
  \item the averaged spectral norm loss:
$
\text{ASL} = \frac{1}{G}\sum_{z \in \mathcal{U}}\sum_{k=1}^{K}\|\hat{\boldsymbol{\Theta}}_{k}^{z} - \boldsymbol{\Theta}_{k}^{z}\|_{2},
$
  \item the averaged Frobenius norm loss:
$
\text{AFL} = \frac{1}{G}\sum_{z \in \mathcal{U}}\sum_{k=1}^{K}\|\hat{\boldsymbol{\Theta}}_{k}^{z} - \boldsymbol{\Theta}_{k}^{z}\|_{\text{F}},
$
  \item the averaged Kullback-Leibler loss:
$
\text{AKL} = \frac{1}{G}\sum_{z \in \mathcal{U}}\sum_{k=1}^{K}\text{KL}(\boldsymbol{\Theta}_{k}^{-1,z}, \hat{\boldsymbol{\Theta}}_{k}^{-1,z}),
$
where $\text{KL}(\boldsymbol{\Theta}^{-1}, \hat{\boldsymbol{\Theta}}^{-1}) = \text{tr}(\boldsymbol{\Theta}^{-1}\hat{\boldsymbol{\Theta}}) - \log|\boldsymbol{\Theta}^{-1}\hat{\boldsymbol{\Theta}}| - p$.
  \item the average squared error (RASE) for estimated mixing proportions:
$$
\text{RASE}_{\pi}^{2} = \frac{1}{G}\sum_{z \in \mathcal{U}}\sum_{k=1}^{K}\left(\hat{\pi}_{k}^{z} - \pi_{k}^{z}\right)^{2}.
$$
  \item the average true positive rate (ATPR) and average false positive rate (AFPR):
$$
\text{ATPR} = \frac{1}{G}\sum_{z \in \mathcal{U}}\frac{1}{K}\sum_{k=1}^{K}\text{TPR}_{k}^{z}
\quad
\text{and}
\quad
\text{AFPR} = \frac{1}{G}\sum_{z \in \mathcal{U}}\frac{1}{K}\sum_{k=1}^{K}\text{FPR}_{k}^{z}.
$$
\end{itemize}

\subsection{Mixture of AR and block diagonal dependencies}


In the first simulation study, we consider the nonparametric mixture of AR(1) $(\boldsymbol{\Sigma}_{1}^{0} = 0.4^{|i-j|})$ and a two-block diagonal structure when $z=0$. To construct $\boldsymbol{\Theta}_{2}^{0}$, when $p=50$ (or $p=100$), we randomly choose $25$ (or $50$) edges from the first and the second block of $\boldsymbol{\Theta}_{2}^{0}$ respectively. The corresponding entries in $\boldsymbol{\Theta}_{2}^{0}$ are uniformly generated from $[-0.2, -0.1]$ and the diagonal elements are set to be $0.25$. We fix the mean vector as zero and generate $50$ samples for each of $11$ equally spaced points. Here we consider growing edges of both graphs: at each of the remaining $10$ equally spaced points, we randomly add $5$ new edges to 1st mixture towards the AR(2) structure, and randomly add $5$ new edges to each block of 2nd mixture when $p=50$. When $p=100$, we add $20$ new edges instead of $5$. Here, $\boldsymbol{\Theta}_{1}^{z}$ and $\boldsymbol{\Theta}_{2}^{z}$ evolves very smoothly. If necessary, we increase diagonal elements to keep precision matrices positive definite.


First of all, we check the performance of choosing number of mixtures. For a given bandwidth $h$, we report the frequencies of $\min_{\lambda}\text{BIC}$ over $100$ repeats in Table \ref{tab:t1}. As shown in Table \ref{tab:t1}, undersmoothing may cause an underestimated $K$. Hence, we should choose number of mixtures over all bandwidths, instead of using $\min_{\lambda}\mbox{BIC}$ for some given bandwidth. Our proposed BIC criterion based on $\min_{\lambda, h}\mbox{BIC}$ has a convincing performance in  choosing correct number of mixtures.


\begin{table}[!htb]
\caption{Frequencies of choosing number of mixtures using $\min_{\lambda,h}\mbox{BIC}$ over 100 repeats.}
\label{tab:t1}
\small{
\begin{center}
\begin{tabular}{c|ccc|ccc}
\hline
& $K=1$ & $K=2$ & $K=3$ & $K=1$ & $K=2$ & $K=3$ \\
\hline
& \multicolumn{3}{c|}{{$p=50$}} & \multicolumn{3}{c}{{$p=100$}} \\
\hline
$h=0.65$ & 82 & 18 & 0 & 91 & 9 & 0 \\
\hline
$h=0.85$ & 16 & 84 & 0 & 3 & 97 & 0 \\
\hline
$h=1.05$ & 2 & 96 & 2 & 0 & 100 & 0 \\
\hline
$h=1.25$ & 1 & 97 & 2 & 1 & 99 & 0 \\
\hline
$h=1.45$ & 1 & 97 & 2 & 0 & 100 & 0 \\
\hline
\hline
$\min_{\lambda, h}\mbox{BIC}$ & 1 & 97 & 2 & 0 & 100 & 0 \\
\hline
\end{tabular}
\end{center}
}
\end{table}

Next, we examine the graphical model selection performance using ATPR and AFPR and the graph and mixing proportions estimation performance using ASL, AFL, AKL and $\text{RASE}_{\pi}$. The results are reported in Tables \ref{tab:t2} and \ref{tab:t3} respectively. Overall, our proposed method achieves a fairly high ATPR and a reasonably low AFPR.  In this simulation, both estimation and selection performances of our proposed method is not sensitive to bandwidths. Since two graphs evolves very smoothly, it tends to give better result with larger bandwidth. However, as shown in $\text{RASE}_{\pi}$, oversmoothing may lead to a slightly biased estimation of mixing proportions.


\begin{table}[!htb]
\caption{Comparison of graphical model selection performance using ATPR and AFPR.}
\label{tab:t2}
\small{
\begin{center}
\begin{tabular}{c|cc|cc}
\hline
& ATPR & AFPR & ATPR & AFPR \\
\hline
& \multicolumn{2}{c|}{{$p=50$}}  & \multicolumn{2}{c}{{$p=100$}} \\
\hline
$h=0.65$ & 0.9258 (0.0113) & 0.1906 (0.0010) & 0.8927 (0.0090) & 0.1072 (0.0006) \\
\hline
$h=0.85$ & 0.9390 (0.0116) & 0.1674 (0.0011) & 0.9030 (0.0103) & 0.0920 (0.0005) \\
\hline
$h=1.05$ & 0.9481 (0.0111) & 0.1561 (0.0011) & 0.9081 (0.0105) & 0.0845 (0.0005) \\
\hline
$h=1.25$ & 0.9523 (0.0107) & 0.1519 (0.0010) & 0.9110 (0.0098) & 0.0810 (0.0006) \\
\hline
$h=1.45$ & 0.9550 (0.0103) & 0.1503 (0.0010) & 0.9136 (0.0094) & 0.0803 (0.0005) \\
\hline
\end{tabular}
\end{center}
}
\end{table}

\begin{table}[!htb]
\caption{Comparison of estimation performance using ASL, AFL, AKL, and $\mbox{RASE}_{\pi}$.}
\label{tab:t3}
\small{
\begin{center}
\begin{tabular}{c|ccccc}
\hline
& ASL & AFL & AKL & $\mbox{RASE}_{\pi}$ \\
\hline
& \multicolumn{4}{c}{{$p=50$}} \\
\hline
$h=0.65$ & 1.9036 (0.0290) & 6.7950 (0.1320) & 10.7644 (0.8102) & 0.1431 (0.0045) \\
\hline
$h=0.85$ & 1.8572 (0.0290) & 6.5677 (0.1348) & 9.5934 (0.8485) & 0.1395 (0.0041) \\
\hline
$h=1.05$ & 1.8442 (0.0279) & 6.4714 (0.1304) & 9.0155 (0.8427) & 0.1416 (0.0038) \\
\hline
$h=1.25$ & 1.8398 (0.0267) & 6.4272 (0.1246) & 8.7179 (0.8077) & 0.1424 (0.0035) \\
\hline
$h=1.45$ & 1.8383 (0.0259) & 6.4055 (0.1205) & 8.5558 (0.7809) & 0.1424 (0.0032) \\
\hline
& \multicolumn{4}{c}{{$p=100$}} \\
\hline
$h=0.65$ & 2.0519 (0.0170) & 10.0884 (0.1156) & 20.5775 (1.0949) & 0.3670 (0.0011) \\
\hline
$h=0.85$ & 2.0186 (0.0170) & 9.9191 (0.1260) & 19.3055 (1.3147) & 0.3649 (0.0011) \\
\hline
$h=1.05$ & 2.0129 (0.0170) & 9.8929 (0.1299) & 18.9758 (1.4092) & 0.3645 (0.0010) \\
\hline
$h=1.25$ & 2.0105 (0.0164) & 9.8738 (0.1250) & 18.6405 (1.3602) & 0.3635 (0.0008) \\
\hline
$h=1.45$ & 2.0091 (0.0159) & 9.8648 (0.1224) & 18.5014 (1.3464) & 0.3643 (0.0011) \\
\hline
\end{tabular}
\end{center}
}
\end{table}


\subsection{Mixture of AR and random sparse dependencies}


In the second simulation study, we consider the nonparametric mixture of AR(1) $(\boldsymbol{\Sigma}_{1}^{0} = 0.4^{|i-j|})$ and a random sparse structure when $z=0$. To construct $\boldsymbol{\Theta}_{2}^{0}$, when $p=50$ (or $p=100$), we set its diagonal elements as $0.25$ and randomly choose $55$ (or $100$) off-diagonal entries to be drawn uniformly from $[-0.25, -0.22]$. We fix the mean vector as zero and generate $50$ samples for each of $11$ equally spaced points. Here we consider simultaneously growing and decaying edges of both graphs: at each equally spaced point, we randomly add $5$ new edges to both graphs and randomly remove $5$ existing edges when $p=50$. When $p=100$, we randomly add $20$ new edges and remove $20$ existing edges. $\boldsymbol{\Theta}_{1}^{z}$ and $\boldsymbol{\Theta}_{2}^{z}$ evolves less smoothly than those in the first simulation.





\begin{table}[!htb]
\caption{Frequencies of choosing number of mixtures using $\min_{\lambda,h}\mbox{BIC}$ over 100 repeats.}
\label{tab:t4}
\small{
\begin{center}
\begin{tabular}{c|ccc|ccc}
\hline
& $K=1$ & $K=2$ & $K=3$ & $K=1$ & $K=2$ & $K=3$ \\
\hline
& \multicolumn{3}{c|}{{$p=50$}} & \multicolumn{3}{c}{{$p=100$}} \\
\hline
$h=0.65$ & 95 & 5 & 0 & 100 & 0 & 0 \\
\hline
$h=0.85$ & 23 & 77 & 0 & 5 & 95 & 0 \\
\hline
$h=1.05$ & 10 & 89 & 1 & 1 & 99 & 0 \\
\hline
$h=1.25$ & 4 & 94 & 2 & 0 & 100 & 0 \\
\hline
$h=1.45$ & 0 & 97 & 3 & 0 & 100 & 0 \\
\hline
\hline
$\min_{\lambda, h}\mbox{BIC}$ & 0 & 97 & 3 & 0 & 100 & 0 \\
\hline
\end{tabular}
\end{center}
}
\end{table}

Table \ref{tab:t4} summarizes the performance of choosing number of mixtures. For a given bandwidth $h$, we report the frequencies of $\min_{\lambda}\text{BIC}$ over $100$ repeats. As shown in Table \ref{tab:t4}, undersmoothing again may cause an underestimated $K$. Our proposed BIC criterion has a consistent performance in choosing the correct number of mixtures.

The selection and the estimation performance is given in Tables \ref{tab:t5}--\ref{tab:t6}. Overall, our proposed method achieves a fairly high ATPR, a reasonably low AFPR, and the appealing estimation losses. Since two graphs evolves less smoothly, oversmoothing may lead to a more biased estimation of precision matrices and mixing proportions.

\begin{table}[!htb]
\caption{Comparison of graphical model selection performance using ATPR and AFPR.}
\label{tab:t5}
\small{
\begin{center}
\begin{tabular}{c|cc|cc}
\hline
& ATPR & AFPR & ATPR & AFPR \\
\hline
& \multicolumn{2}{c|}{{$p=50$}}  & \multicolumn{2}{c}{{$p=100$}} \\
\hline
$h=0.65$ & 0.8732 (0.0160) & 0.1769 (0.0011) & 0.8890 (0.0113) & 0.1095 (0.0005) \\
\hline
$h=0.85$ & 0.8859 (0.0177) & 0.1555 (0.0013) & 0.9042 (0.0118) & 0.0923 (0.0005) \\
\hline
$h=1.05$ & 0.8889 (0.0183) & 0.1442 (0.0014) & 0.9098 (0.0118) & 0.0836 (0.0006) \\
\hline
$h=1.25$ & 0.8888 (0.0184) & 0.1398 (0.0014) & 0.9117 (0.0115) & 0.0803 (0.0006) \\
\hline
$h=1.45$ & 0.8894 (0.0185) & 0.1382 (0.0014) & 0.9124 (0.0113) & 0.0793 (0.0006) \\
\hline
\end{tabular}
\end{center}
}
\end{table}

\begin{table}[!htb]
\caption{Comparison of estimation performance using ASL, AFL, AKL, and $\mbox{RASE}_{\pi}$.}
\label{tab:t6}
\small{
\begin{center}
\begin{tabular}{c|ccccc}
\hline
& ASL & AFL & AKL & $\mbox{RASE}_{\pi}$ \\
\hline
& \multicolumn{4}{c}{{$p=50$}} \\
\hline
$h=0.65$ & 2.1992 (0.0352) & 7.4927 (0.1548) & 12.7527 (0.9811) & 0.1589 (0.0055) \\
\hline
$h=0.85$ & 2.1312 (0.0361) & 7.2900 (0.1646) & 11.8158 (1.0692) & 0.1584 (0.0055) \\
\hline
$h=1.05$ & 2.1031 (0.0360) & 7.2195 (0.1680) & 11.5378 (1.1045) & 0.1577 (0.0052) \\
\hline
$h=1.25$ & 2.0973 (0.0359) & 7.2041 (0.1683) & 11.4536 (1.1069) & 0.1577 (0.0050) \\
\hline
$h=1.45$ & 2.0965 (0.0357) & 7.2018 (0.1682) & 11.4282 (1.1064) & 0.1579 (0.0049) \\
\hline
& \multicolumn{4}{c}{{$p=100$}} \\
\hline
$h=0.65$ & 2.2518 (0.0201) & 10.0592 (0.1270) & 20.0166 (1.1550) & 0.3684 (0.0012) \\
\hline
$h=0.85$ & 2.2053 (0.0186) & 9.8330 (0.1261) & 18.2240 (1.2288) & 0.3650 (0.0011) \\
\hline
$h=1.05$ & 2.1981 (0.0174) & 9.7937 (0.1237) & 17.7696 (1.2573) & 0.3635 (0.0009) \\
\hline
$h=1.25$ & 2.2014 (0.0169) & 9.7985 (0.1221) & 17.6933 (1.2571) & 0.3633 (0.0008) \\
\hline
$h=1.45$ & 2.2039 (0.0166) & 9.8037 (0.1209) & 17.6861 (1.2501) & 0.3636 (0.0008) \\
\hline
\end{tabular}
\end{center}
}
\end{table}

\section{\large Application to ADHD Imaging Data}

This section applies our proposed method to estimate time varying brain functional connectivity from ADHD-200 Global Competition data \citep{biswal-etal-2010}. The ADHD-200 training dataset has fMRI images, diagnosis information and other demographic variables (e.g., Age, IQ, gender, handedness) of $776$ subjects from 8 different sites. We focus on $N=284$ subjects whose ages range from 9 to 13, since observed ages are not uniformly distributed outside the range 9-13. Each fMRI image has measurements for $p=264$ seed voxels.
Among these subjects, 186 subjects are typically developing controls and 98 subjects are diagnosed with ADHD. We choose Age as the covariate $Z$, and normalize it to $[0, 1]$. We consider three interested ages: 9, 11 and 13.

To estimate the nonparametric mixture, we first determine the number of mixtures by using our proposed BIC in Section 3.2. BIC is minimized when $K=2$, which implies that there are two heterogenous groups with certain commonality. After fixing $K=2$, we solve two heterogeneous graphical models at the aforementioned three interested ages respectively.
Table \ref{tab:t10} shows the estimated mixing proportions. We can see a slightly increasing trend in first mixture proportion as age increases. In what follows, we will investigate these two subpopulations through profiling their demographic variables, and explain their differences from three different perspectives: site information, impact of gender, and impact of IQ. Our results are supported by existing scientific findings.


\begin{table}[!htb]
\caption{Estimated mixing proportions at the Age 9, Age 11 and Age 13.}
\label{tab:t10}
\small{
\begin{center}
        \begin{tabular}{c|ccc}
        \hline
            & Age 9 & Age 11 & Age 13 \\
        \hline
            $\hat{\pi}_{1}$ & 0.6742 & 0.6870 & 0.7011 \\
        \hline
            $\hat{\pi}_{2}$ & 0.3258 & 0.3130 & 0.2989 \\
        \hline
        \end{tabular}
\end{center}
}
\end{table}



\begin{figure}[!htb]
\caption{Site information: 1st mixture (left panel) and 2nd mixture (right panel).}
\label{fig:Pie}
\vspace{-4.5em}
\centering
\begin{subfigure}{.4\textwidth}
\centering
  \includegraphics[width=1.05\linewidth]{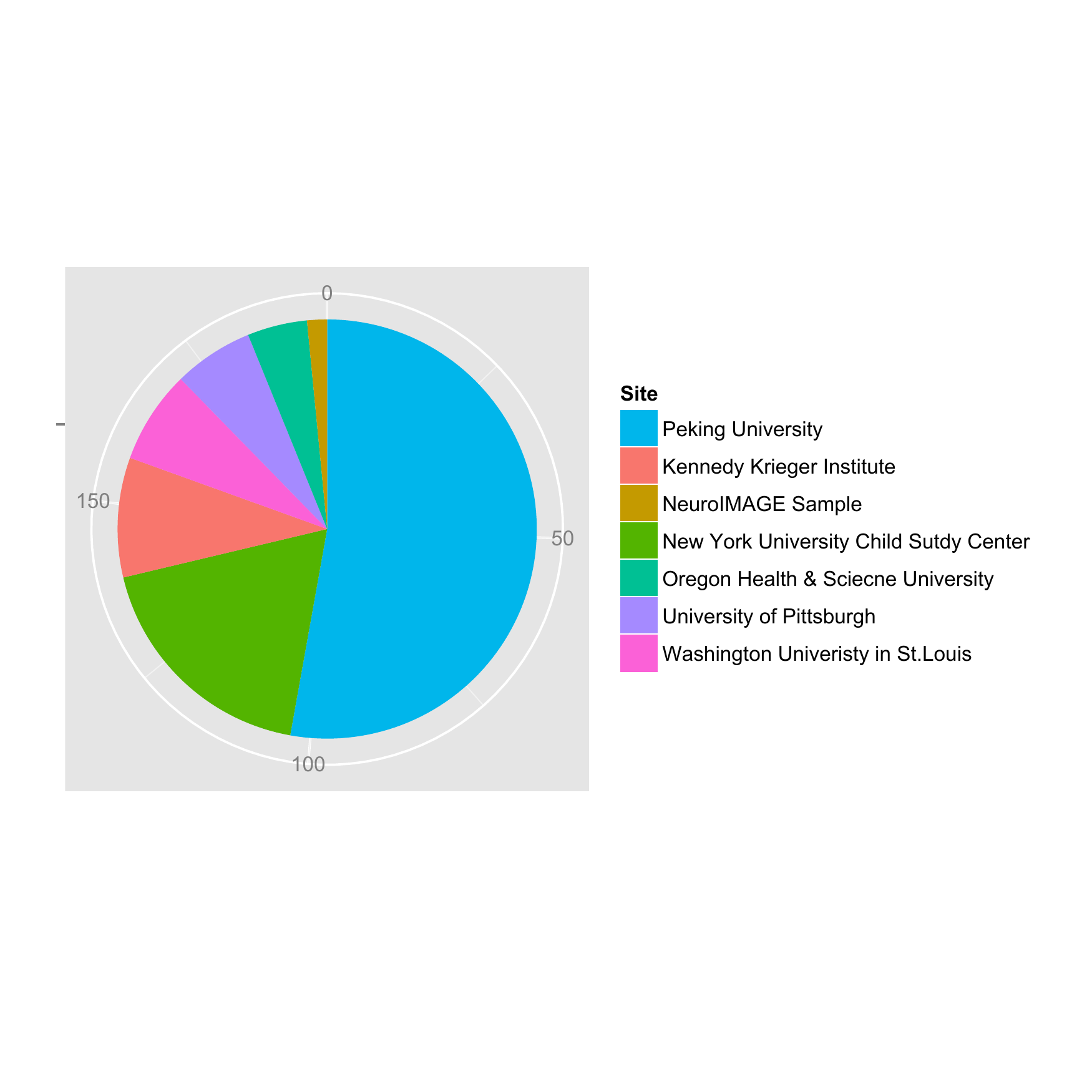}
\end{subfigure}%
\begin{subfigure}{.4\textwidth}
  \centering
  \includegraphics[width=1.05\linewidth]{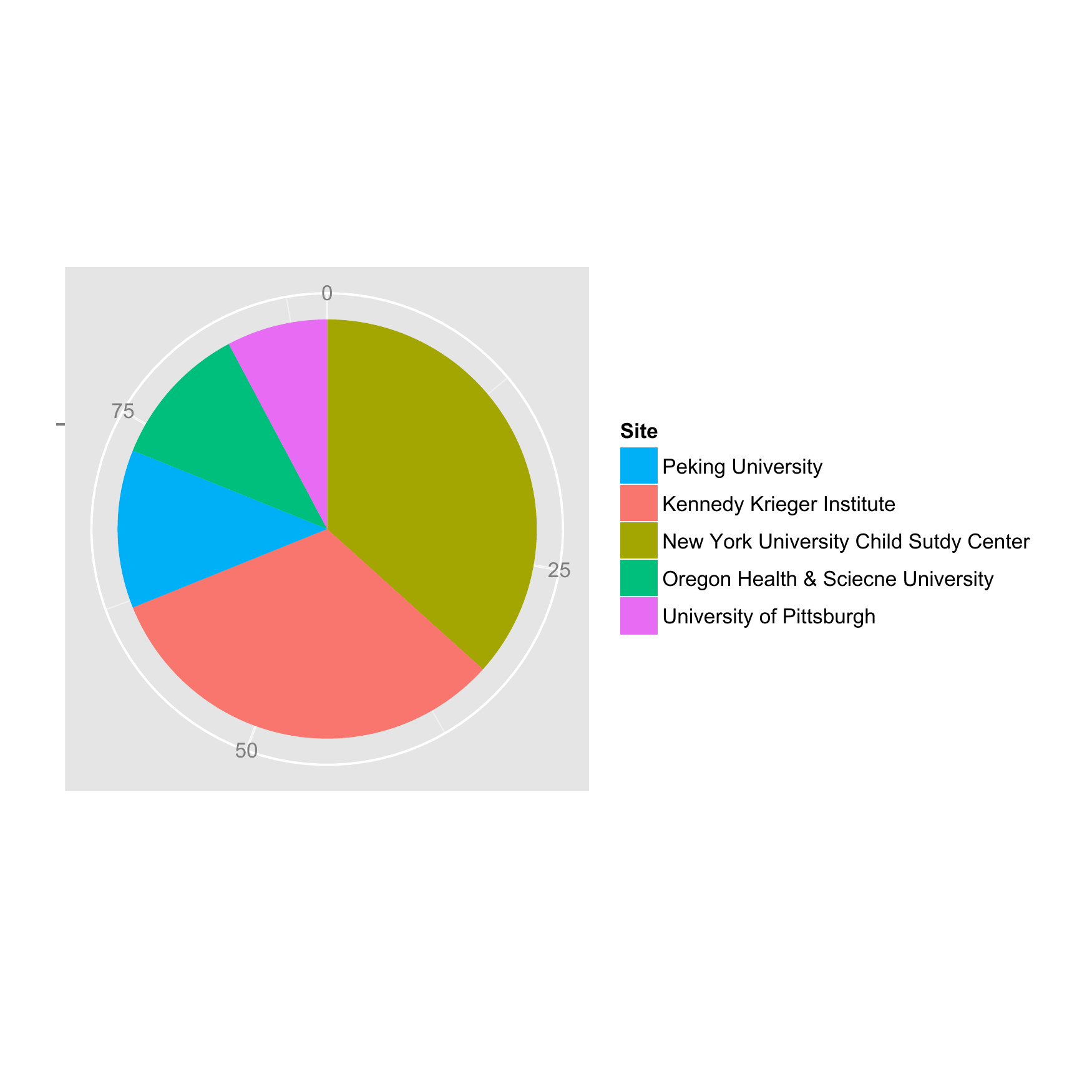}
\end{subfigure}
\vspace{-4.5em}
\end{figure}

Firstly, we explore how subpopulations are composed with subjects from various sites. As shown in Figure \ref{fig:Pie},
both mixtures are formed with subjects from heterogeneous locations. This confirms the previous study that geographic locations are not sufficient to explain variability of ADHD \citep{polanczyk-etal-2007}. Hence we need to further investigate their individual commonality beyond geographic locations.

Secondly, we study the relationship between gender and ADHD status (i.e., whether the subject has ADHD or not). In the existing literature, gender difference in ADHD is an important research topic  \citep{arnold-1996}. There are two main arguments about the reason why the boys are more likely to be diagnosed with ADHD than girls. On one hand, some researchers insist that there is actually gender difference in ADHD. On the other hand, other studies show that there are other issues for example, other demographic covariates or the bias of the diagnose test that affects diagnose results. Here, we test the independence between gender and ADHD status in two mixture respectively. Their corresponding contingency tables are given in Table \ref{tab:t11}. For the first mixture, we compute the chi-square test statistic, $\chi^{2} = 7.3106$ with associated $\mbox{p-value} = 0.0069$. It implies that there is an indeed relationship between gender and ADHD status in this mixture. While for the second mixture, we have chi-square test statistic, $\chi^{2} = 0.5237$ with associate $\mbox{p-value} = 0.4693$, which indicates that gender and ADHD status could be independent in the second mixture. Therefore, we successfully identify two heterogeneous subpopulations: the first one is consistent with the previous study that ADHD is diagnosed at a significantly higher rate in boys than in girls \citep{sauver-etal-2004}; the second one is consistent with another study that there may exist other covariates strongly related to the ADHD, but which are not different for boys and girls so that ADHD are not systematically different for boys and girls \citep{bauermeister-etal-2007}. Both subpopulations support their corresponding scientific findings respectively.

\begin{table}[!htb]
\caption{Contingency tables with gender and whether they have ADHD or not.} \label{tab:t11}
\vspace{-.2in}
\small{
\begin{center}
\begin{tabular}{cc}
    \begin{minipage}{.5\linewidth}
    \centering{(A) 1st mixture}
        \begin{tabular}{c|c|c|c}
        \hline
            & Male & Female & Total \\
        \hline
            Control & 77 & 57 & 134 \\
        \hline
             ADHD & 48 & 13 & 61 \\
        \hline
             Total & 125 & 70 & 195 \\
        \hline
        \end{tabular}
    \end{minipage} &

    \begin{minipage}{.5\linewidth}
    \centering{(B) 2nd mixture}
    \medskip
    \begin{tabular}{c|c|c|c}
        \hline
            & Male & Female & Total \\
        \hline
            Control & 30 & 22 & 52 \\
        \hline
             ADHD & 25 & 12 & 37 \\
        \hline
             Total & 55 & 34 & 89 \\
        \hline
        \end{tabular}
    \end{minipage}
\end{tabular}
\end{center}
}
\end{table}


Thirdly, we investigate the full scale IQ scores for both mixtures. Table \ref{tab:t9} summarizes the average full scale IQ for each mixture. For the first mixture, we can clearly see typically developing children have a higher average IQ score than the children with ADHD: female control's average IQ is 16 higher than female ADHD's average IQ, and male control's average IQ is 12.32 higher than male ADHD's average IQ. This result shows that full scale IQ scores are reliably different between individuals with ADHD and typically developing controls in the first mixture, which is consistent with \cite{Frazier-etal-2004,garcia-etal-1997}. However, the second mixture exhibits a very different aspect: typically developing children in the second mixture have a lower average IQ score compared to those in the first mixture, while ADHD children in the second mixture have a higher average IQ score than those in the first mixture. Therefore, there is only no significant difference in terms of average IQ between typically growing children and children with ADHD in the second mixture: female control's average IQ is 7.77 higher than female ADHD's average IQ, and male control's average IQ is 4.30 higher than male ADHD's average IQ. Based on the full scale IQ, these two heterogeneous mixtures further justifies the power of our proposed method.

\begin{table}[!htb]
\caption{Average full scale IQ scores for both mixtures.} \label{tab:t9}
\small{
\begin{center}
        \begin{tabular}{c|cc}
        \hline
            & First & Second \\
        \hline
            Male Control  & 118.85 & 112.30 \\
        \hline
            Female Control & 115.54 & 111.77 \\
        \hline
            Male ADHD & 106.53 & 108.00 \\
        \hline
            Female ADHD & 99.54 & 104.00 \\
        \hline
        \end{tabular}
\end{center}
}
\end{table}

\begin{singlespace}
\bibliographystyle{agsm}
\bibliography{graph}
\end{singlespace}

\end{document}